\documentstyle[12pt,twoside,fleqn,espcrc1,psfig]{article}

\newcommand{\AmS}{{\protect\the\textfont2
  A\kern-.1667em\lower.5ex\hbox{M}\kern-.125emS}}

\title{Hydrodynamics and collective behaviour in relativistic nuclear
collisions\thanks{This work was supported by the Director, Office of Energy
        Research, Division of Nuclear Physics of the Office of 
        High Energy and Nuclear Physics of the U.S.\ Department of 
        Energy under Contract No.\ DE-FG-02-93ER-40764.}}

\author{Dirk H.\ Rischke \address{
        Physics Department, Pupin
        Physics Laboratories, Columbia University, \\ 
        538W 120th Street, New York, NY 10027, U.S.A.}}

\begin{document}
\maketitle

\begin{abstract}
Hydrodynamics is applied to describe 
the dynamics of relativistic heavy-ion collisions. The
focus of the present study is the influence of a possible (phase) transition 
to the quark--gluon plasma in the nuclear matter equation of state 
on collective observables, such as the lifetime of the system and
the transverse directed flow of matter. It is shown that such a transition
leads to a softening of the equation of state, and consequently to
a time-delayed expansion which is in principle observable
via two--particle correlation functions.
Moreover, the delayed expansion leads to a local minimum in the
excitation function of transverse directed flow around AGS energies.
\end{abstract}

\section{INTRODUCTION}

Lattice calculations of the thermodynamical functions of
quantum chromodynamics (QCD) indicate \cite{QM96} 
that, (at zero net baryon density) in the vicinity of a 
critical temperature $T_c \sim 160$ MeV, strongly interacting
matter undergoes a rapid transition from a (chirally broken, confined) 
hadronic phase to a (chirally symmetric, deconfined) quark--gluon plasma
(QGP). The width $\Delta T$ of that transition region is presently only known 
to be in the range $0 \leq \Delta T < 0.1 \, T_c \sim 16$ MeV.
Therefore, one cannot yet conclude whether the transition is a first 
order phase transition ($\Delta T=0$), or merely a rapid increase 
of the entropy density associated with the change from $d_H$ hadronic to
$d_Q$ quark and gluon degrees of freedom. 

One of the primary goals of present relativistic heavy-ion physics 
is the creation and experimental observation of the predicted
QGP phase of matter. Many signatures have been proposed
such as electromagnetic radiation of thermal dileptons and 
photons \cite{dilepton}, $J/\Psi$--suppression \cite{JPsi}, 
jet quenching \cite{jetquench}, strangelet formation \cite{strangelet}, or
disordered chiral condensates (DCC's) \cite{wilcek}. 
These signatures, however, do not depend directly on the actual 
form of the nuclear matter equation of state (EoS).
Thermal electromagnetic radiation is, for instance, generic to any
hot system, independent from its degrees of freedom (as long
as they have electromagnetic charge).
For example, it was shown \cite{kapusta} that a hot hadron
gas shines as brightly as a QGP. Similarly, jet quenching and
$J/\Psi$--suppression \cite{footnote1}
are generic consequences of final state interactions
in any form of dense matter \cite{huefner}. 
Finally, strangelet or DCC formation require very specific
assumptions about the dynamical evolution of the system.

It is therefore of interest to study signals that are more
directly related to the QCD equation of state. 
Signals of this type emerge from the influence of the
EoS on the {\em collective\/} dynamical evolution of 
the system. Relativistic hydrodynamics \cite{strottman} 
is the most suitable approach to study these signals, since it is the only 
dynamical model which provides a {\em direct\/} link between collective 
observables and the EoS. 
Of course, the use of such an approach requires a strong dynamical
assumption, namely that the equilibration rates are much larger
than typical gradients of thermodynamic quantities in the system. 
At least for high energy density QCD matter, radiative gluon energy loss was
estimated to be sufficiently large \cite{eloss} to support local equilibration
on time scales less than 1 fm/c. In the following, I
therefore neglect dissipative effects and assume the validity of 
ideal hydrodynamics to compute the collective evolution of the system.

It was shown \cite{shuryak,test1,dhrmg,dhrmg2} 
that the transition to the QGP {\em softens\/} the EoS
in the transition region, and thus reduces the tendency
of matter to expand on account of its internal pressure.
This, in turn, delays the expansion and considerably
prolongs the lifetime of the system. It was moreover shown \cite{dhrmg2} 
that this prolongation of the lifetime (as compared to the expansion of an
ideal gas without transition) is in
principle observable via an enhancement of the ratio of
inverse widths, $R_{\rm \, out}/R_{\rm \,side}$,
of the two--particle correlation function in out-- and
side--direction. (This signal was originally proposed by Pratt and 
Bertsch \cite{pratt}.) Another aspect \cite{csernai,puersuen} of the 
delayed expansion is the reduction of the 
transverse directed flow in semi-peripheral collisions that can be 
readily tested experimentally at fixed target energies \cite{E877}. 

In this paper I summarize the essential physics 
of the softening of the EoS in the transition region, and 
discuss as observable consequences
the time-delayed expansion and the subsequent enhancement of
$R_{\rm \,out}/R_{\rm \, side}$, and the disappearance of the transverse
directed flow. Natural units $\hbar = c = k_B = 1$ are used 
throughout this paper.

\section{THE QCD PHASE TRANSITION AND SOFTENING OF THE EQUATION OF STATE}

Available lattice data for the entropy density in full QCD
can be approximated by the simple parametrization \cite{dhrmg,dhrmg2,blaizot}
\begin{equation} \label{eos}
\frac{s}{s_c}(T) = \left[\frac{T}{T_c}\right]^3 
\left( 1 + \frac{d_Q-d_H}{d_Q+d_H} \, 
\tanh \left[ \frac{T-T_c}{\Delta T} \right] \right)~,
\end{equation}
where $s_c = const. \times (d_Q+d_H)\, T_c^3$ 
is the entropy density at $T_c$.
Pressure $p$ and energy density $\epsilon$ follow
then from  thermodynamical relationships.
For $\Delta T=0$, the EoS (\ref{eos}) reduces
to the MIT bag EoS \cite{MIT} with bag constant
$B= \frac{1}{2}\, (d_Q/d_H-1)\, T_c\, s_c /\,(d_Q/d_H+1)$.
If one measures energies in units of $T_c$ and
energy densities in units of the
enthalpy density $\epsilon_c +p_c =
T_c\, s_c$, the EoS (\ref{eos})
depends only on the ratio $d_Q/d_H$, and not on $d_Q$ and $d_H$
separately. For $\Delta T =0$, this ratio determines the
latent heat (density) $\epsilon_Q - \epsilon_H \equiv 4 B$.
(Here, $\epsilon_Q = \frac{1}{2}\, (4\, d_Q/d_H -1)\, T_c\, s_c\,
/\,(d_Q/d_H+1)$ is the energy density at the phase boundary between
mixed phase and QGP, $\epsilon_H = \frac{3}{2}\, T_c\, s_c\,
/\,(d_Q/d_H+1)$ is that at the boundary between mixed and hadronic phase.)

For the case $d_H =3$ (corresponding to an ultrarelativistic gas 
of pions) and $d_Q = 37$ (corresponding to two massless flavours
of quarks and antiquarks, and eight massless gluons), 
the latent heat, $4B=1.7\, T_c\, s_c \simeq 1.272 $ GeV fm$^{-3}$, is large.
On the other hand, including a resonance gas 
in the hadronic phase and/or reducing
the effective number of degrees of freedom
on the QGP side \cite{neumann}, $d_Q/d_H = 3$ may be 
taken as a (perhaps more realistic) lower limit, with a smaller latent heat 
$\epsilon_Q - \epsilon_H = T_c\, s_c$.
Assuming that the high-temperature phase consists of gluons
only (such as expected for the ``hot-glue scenario''
\cite{glue}) this would then
correspond to about $400$ MeV fm$^{-3}$ in physical units.
In the following, I shall focus on the case $d_Q/d_H=37/3$, for
a discussion of $d_Q/d_H = 3$, I refer to \cite{dhrmg2}.

\begin{figure} \hspace*{0.3in} 
\psfig{figure=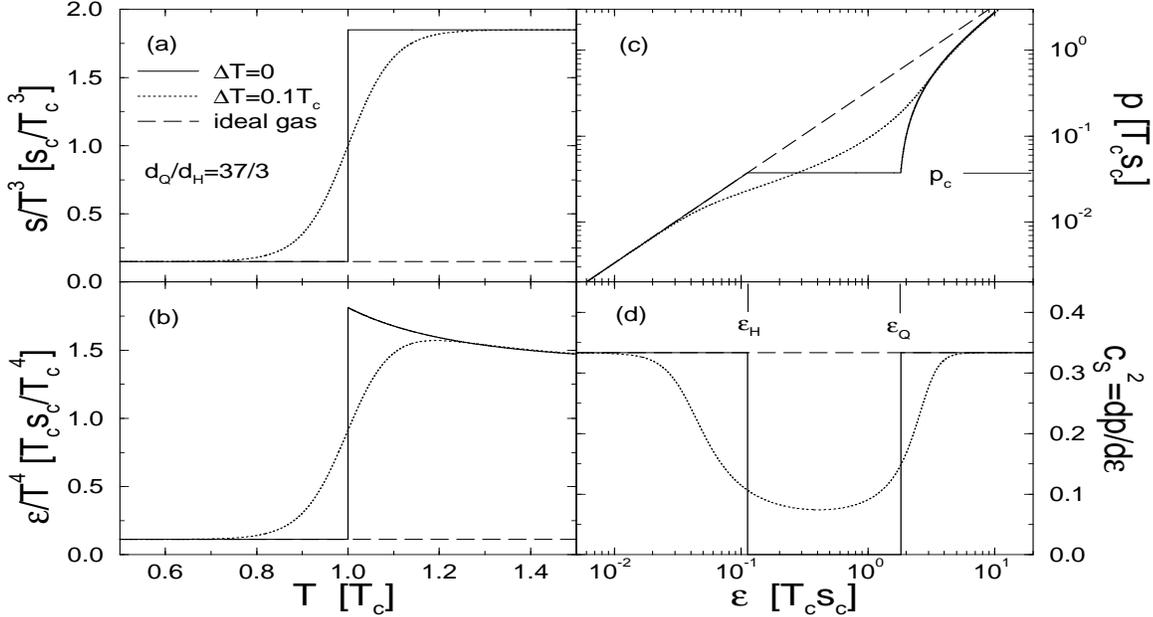,width=3.5in,height=4.5in,angle=-90}
\vspace*{-1.5cm}
\caption{The equation of state: (a) entropy density divided by $T^3$,
(b) energy density divided by $T^4$ as functions of temperature, (c)
pressure, and (d) the velocity of sound squared as functions of
energy density. Solid lines are for a first order transition ($\Delta
T=0$), dotted lines for a smooth transition with $\Delta T= 0.1\, T_c$,
and dashed lines are for an ideal gas with $d_H$ degrees of freedom.
The ratio of degrees of freedom is $d_Q/d_H=37/3$.}
\end{figure}

Fig.\ 1 shows (a) the entropy density and (b) the energy density 
as functions of temperature, and (c) the pressure 
and (d) the velocity of sound squared $c_S^2 \equiv {\rm d}p/
{\rm d}\epsilon$ as functions of energy density
for $\Delta T=0$, $0.1\, T_c$, and an ideal gas with $d_H$ degrees
of freedom for $d_Q/d_H = 37/3$. 
Figs.\ 1 (a,b) present the thermodynamic functions in
a form to facilitate comparison with lattice data. 
Present lattice data for full QCD can be
approximated with a choice of $\Delta T$ in 
the range $0 \leq \Delta T < 0.1\, T_c$. In the
hydrodynamical context, however, Figs.\ 1 (c,d) are more relevant. 
As can be seen in (c), for $\Delta T=0$ the pressure stays constant
in the mixed phase $\epsilon_H \leq \epsilon \leq \epsilon_Q$.
Hydrodynamical expansion is, however, driven by 
pressure {\em gradients}. It is therefore (the square of) the velocity 
of sound $c_S^2 = {\rm d}p/{\rm d}\epsilon$, Fig.\ 1 (d), 
that is the most relevant measure of 
the system's tendency to expand. It represents the capability
to perform mechanical work (which is proportional to pressure
gradients ${\rm d}p$) for a given gradient in energy
density ${\rm d} \epsilon$. For $\Delta T =0$,
the velocity of sound vanishes in the mixed phase, i.e.,
mixed phase matter does not expand at all on its own account, 
even if there are strong gradients in the energy density.
This has the consequence that it does not perform 
mechanical work and therefore cools less rapidly.
For finite $\Delta T$, pressure gradients are finite, but still
smaller than for an ideal gas EoS, and
therefore the system's tendency to expand is also reduced, cf.\ Fig.\ 1 (d). 

The reduction of $c_S^2$ in the transition region is commonly
referred to as {\em ``softening''\/} of the EoS, the respective region
of energy densities is called {\em ``soft region''} 
\cite{shuryak,test1,dhrmg,dhrmg2}.
For matter passing through that region during the
expansion phase, the flow will temporarily slow down
or even possibly stall under suitable conditions and
consequently lead to a {\em time delay\/} in the expansion of the system.

\section{HYDRODYNAMICS}

Hydrodynamics is defined by local energy--momentum conservation,
\begin{equation} \label{eom}
\partial_{\mu} T^{\mu \nu} = 0~.
\end{equation}
Under the assumption of local thermodynamical equilibrium (the so-called
``ideal fluid'' approximation) the energy--momentum tensor
$T^{\mu \nu}$ assumes the particularly simple form \cite{LL}
\begin{equation} \label{tmunu}
T^{\mu \nu} = (\epsilon + p)\, u^{\mu} u^{\nu} - p\, 
g^{\mu \nu}~,
\end{equation}
where $u^{\mu} = \gamma\, (1,{\bf v})$ is the 4--velocity
of the fluid (${\bf v}$ is the 3--velocity, 
$\gamma \equiv (1-{\bf v}^2)^{-1/2}$,
$u_{\mu} u^{\mu} = 1$), and $g^{\mu \nu} = {\rm diag}(+,-,-,-)$
is the metric tensor. The system of equations (\ref{eom}) is closed 
by choosing an EoS
in the form $p=p(\epsilon)$, i.e., as depicted in Fig.\ 1 (c).
In the ideal fluid approximation, the (equilibrium) EoS
is the {\em only\/} input to the hydrodynamical 
equations of motion (\ref{eom}) that relates to properties of
the matter under consideration and is thus able to influence
the dynamical evolution of the system. The final results are
uniquely determined once a particular initial condition and a 
decoupling (``freeze-out'') hypersurface are specified.

For finite baryon density, one has to also take into account
local conservation of baryon number,
\begin{equation}
\partial_{\mu} N^{\mu} = 0\,\, ,
\end{equation}
where $N^{\mu}= n \, u^{\mu}$ is the baryon 4--current (in the ideal
fluid approximation), $n$ is the baryon density in the local rest frame
of a fluid element. In this case, the EoS has in general
to be provided in the form $p=p(\epsilon,n)$ (see
Fig.\ 1 of Ref.\ \cite{puersuen} for an explicit example).

Numerical methods to solve the hydrodynamical equations have been
discussed for instance in \cite{test1}. In the following, I shall
first discuss the time delay 
in the framework of the simple Landau model \cite{landau}, describing
the one--dimensional expansion of a slab of matter (with infinite
extension in transverse direction). I
then consider time delay in the framework of the so-called Bjorken 
model \cite{bjorken}, which is the three--dimensional expansion of a 
cylinder with boost-invariant initial conditions along its axis. This model
is supposedly a good description of the expansion stage in
ultrarelativistic heavy-ion collisions. Finally, I discuss
the transverse directed flow for semi-peripheral Au+Au collisions. 

The Landau model requires to solve the hydrodynamical equations in time
and only one space direction. Moreover, for the Bjorken model the
cylindrical symmetry and longitudinal boost invariance
reduce the equations of motion to a similar, effectively 1+1 dimensional 
set of equations. To solve them I employ
the relativistic Harten--Lax--van Leer--Einfeldt algorithm \cite{schneider}
tested in \cite{test1} and modified by a Sod predictor--corrector step to 
account for geometry and boost invariance in \cite{dhrmg2}. 
The fully 3+1 dimensional problem
of a semi-peripheral Au+Au collision
is solved via operator splitting and the well-established
SHASTA algorithm \cite{test1,book}.

\begin{figure} \hspace*{0.6in}
\psfig{figure=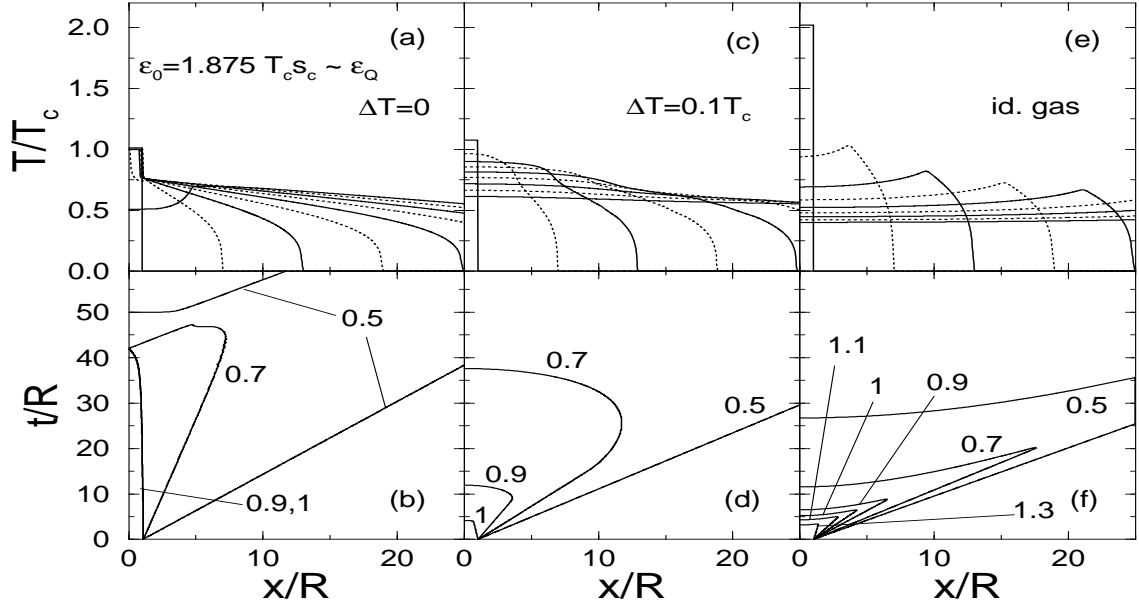,width=3.5in,height=3.9in,angle=-90}
\vspace*{-1.8cm}
\caption{Longitudinal expansion in the Landau model: (a,c,e) temperature
profiles for times $t = 6\, n \lambda R,\, \lambda = 0.99, \, 
n = 0,1,2,...8$ (alternatingly shown as full and dotted lines), 
(b,d,f) isotherms in the space--time diagram. (a,b)
are for $\Delta T=0$, (c,d) for $\Delta T=0.1\, T_c$, and (e,f)
for an ideal gas. The isotherms are labelled by the corresponding
temperature in units of $T_c$.}
\end{figure}

\section{DELAYED EXPANSION AND TWO--PARTICLE CORRELATIONS}

In this section I discuss the delayed expansion and 
observable consequences in the Landau \cite{landau} and
Bjorken model \cite{bjorken} and for the EoS (\ref{eos}). 
Fig.\ 2 presents the hydrodynamic solutions for the
purely 1+1 dimensional expansion in the Landau model, for
an initial (homogeneously distributed)
energy density $\epsilon_0 = 1.875\, T_c\, s_c
\sim \epsilon_Q \simeq 1.4$ GeV fm$^{-3}$. Figs.\ 2 (a,c,e) show
temperature profiles as a function of $x$ (in units of half the
initial extension of the slab, $R$), (b,d,f) isotherms in
the $t-x$ plane (labelled with the corresponding temperatures
in units of $T_c$). As one observes, for a phase transition
with a sharp first order transition ($\Delta T = 0$), Figs.\
2 (a,b), or for a smooth transition ($\Delta T = 0.1\, T_c$), Figs.\
2 (c,d), the system stays hot (i.e., at temperatures around $0.7\, T_c$)
for a rather long time. The reason is, as explained in the preceding section, 
the softening of the EoS in the mixed phase which 
considerably reduces pressure gradients that, for instance, 
drive the expansion in the ideal gas case, Figs.\ 2 (e,f). 

In addition,
for $\Delta T=0$, the type of hydrodynamical expansion solution
changes from an ordinary rarefaction wave to a rarefaction shock
wave \cite{test1,dhrmg}, or so-called deflagration. That deflagration 
has a rather small propagation velocity for energy 
densities around $\epsilon_Q$ \cite{vanH} and thus additionally
prolongs the lifetime of the system. In this case, temperatures as high
as $T_c$ persist for times as long as $42\, R$. In the case $\Delta T=0.1\,
T_c$, such high temperatures vanish almost as fast as in the ideal gas
case, but the softening of the EoS leads to a prolongation
of the lifetime of temperatures around $0.7\, T_c$ as compared to the
ideal gas case. Thus, in view of the uncertainty in the QCD equation of
state, it is rather unlikely that, as discussed in \cite{shuryak},
enhanced electromagnetic radiation
from a long-lived mixed phase is a viable signature for the transition
to the QGP. On the other hand, if the
system freezes out at sufficiently cool temperatures, the long lifetime
of matter with $T \simeq 0.7\, T_c$ could be observed via
two--particle correlations. This idea shall be pursued further in the
following discussion of the more realistic Bjorken expansion scenario.

\begin{figure} \hspace*{0.6in}
\psfig{figure=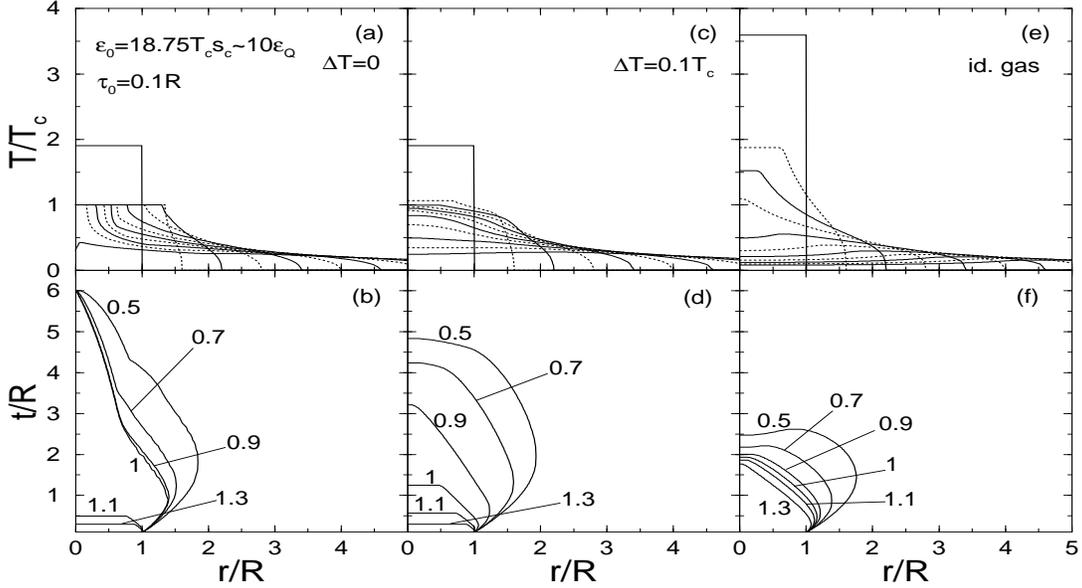,width=3.3in,height=4in,angle=-90}
\vspace*{-3cm}
\caption{Transverse expansion in the Bjorken model. The presentation
is analogous to Fig.\ 2, the initial energy density, however, is
now ten times higher, $\epsilon_0 \sim 10 \epsilon_Q$. The temperature
profiles in (a,c,e) are shown for times $t = \tau_0 + 0.6\, n \lambda
R$, $\tau_0 = 0.1\, R,\, \lambda = 0.99$.}
\end{figure}
 
The main assumption of Bjorken's model
is longitudinal boost invariance which implies that the longitudinal flow
velocity of matter is always given by $v^z \equiv z/t$ \cite{bjorken}. 
The initial conditions are specified at constant proper time
$\tau \equiv \sqrt{t^2-z^2}$. I fix $\tau_0 = 0.1 \, R$, motivated
by the fact that for Au+Au collisions at RHIC energies
equilibration is expected \cite{eloss} 
to occur after $0.5$ fm, while the initial radius $R$ of the hot zone 
is on the order of $5$ fm.
Fig.\ 3 shows hydrodynamic solutions for the (cylindrically symmetric)
transverse expansion of a ``Bjorken cylinder'' (at $z=0$), for
an initial energy density $\epsilon_0 = 18.75\, T_c\, s_c
\sim 10\, \epsilon_Q \simeq 14$ GeV fm$^{-3}$. This value
is expected to be reached through mini-jet production at RHIC 
energies \cite{glue}. Again, for a transition to the QGP the system spends
considerable time in the ``soft region''
of the EoS (corresponding to temperatures around $T_c$), where pressure
gradients are small, and therefore
the expansion is delayed, Figs.\ 3 (a--d), in
comparison to the ideal gas case, Figs.\ 3 (e,f). Note that this
delay now occurs at about $10$ times higher energy density than in
the Landau model. This is due to the fact that the strong 
dilution due to the longitudinal velocity
field has to be compensated so that the
system stays long enough in the ``soft'' transition region.
 
\begin{figure}\hspace*{0.3in} 
\psfig{figure=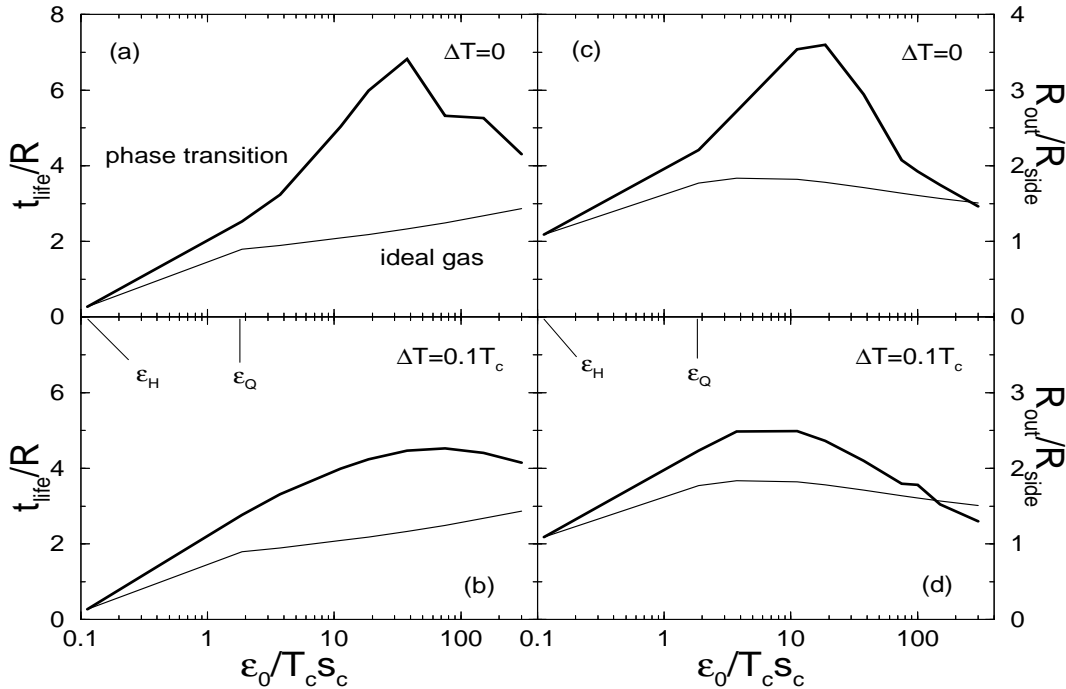,width=3.5in,height=4in,angle=-90}
\vspace*{-1.5cm}
\caption{Lifetimes (a,b) and ratio of inverse widths of correlation functions
(c,d) as a function of initial energy density for the Bjorken expansion. 
The thick lines correspond to the
case of a transition to the QGP, the thin lines to the ideal gas
expansion.}
\end{figure}

Figs.\ 4 (a,b) show the lifetime \cite{footnote2} of matter
with temperature $T=0.7\, T_c$
as a function of initial energy density $\epsilon_0$ for the
Bjorken expansion. One observes a distinguished maximum in the lifetime
associated with the transition to the QGP. In accordance with
the explanation provided above, this maximum of the lifetime occurs at
energy densities {\em above\/} the ``soft region'' of the EoS. 
For systems with zero initial velocity, such as 
a spherically symmetric fireball at rest,
the maximum would occur around $\epsilon_0 \sim \epsilon_Q$ 
\cite{dhrmg,dhrmg2}.

Let us assume that the system decouples at $T=0.7\, T_c$. Then, the
isotherm with $T=0.7\, T_c$ in Figs.\ 3 (b,d,f) is the decoupling or
``freeze-out'' hypersurface. Once this hypersurface is determined, one 
can also calculate the
corresponding particle spectra. For single inclusive spectra,
one commonly employs the formula of Cooper and Frye \cite{cooper}, and
for two--particle correlation functions a suitable generalization
\cite{schlei}. In this manner, one can determine the two--pion
correlation function $C_2(q_{\rm side}, q_{\rm out}, {\bf K})$
as a function of the relative pion momenta
$q_{\rm out}$ or $q_{\rm side}$ for fixed average momentum ${\bf K}$.

For fixed ${\bf K}$
one defines the so-called ``side'' correlation function
as $C_{2,{\rm side}}(q_{\rm side})$ $\equiv C_2(q_{\rm side},0, {\bf K})$, and
the ``out'' correlation function as 
$C_{2,{\rm out}}(q_{\rm out})$ $\equiv C_2(0,q_{\rm out}, {\bf K})$.
The inverse width  of the ``out'' correlation function is a measure
for the duration of particle emission, i.e., the lifetime of the system,
while the inverse width of the ``side'' correlation function measures
its transverse size \cite{pratt}.

Fig.\ 5 shows these functions for (a,b) the case of a sharp first
order transition $\Delta T=0$, (c,d) a smooth transition  $\Delta T=0.1\, T_c$,
and (e,f) the ideal gas expansion. The correlation functions are calculated
along the $T=0.7\, T_c$ isotherms of Figs.\ 3 (b,d,f) with ${\bf K}=(K,0,0)$,
$K = 300$ MeV, and $R = 5$ fm to fix the $q$--scale. The pion mass
is assumed to be $m_\pi = 138$ MeV, and $T_c = 160$ MeV.
Note that the long lifetime of the system in Fig.\ 3 (b) is reflected
in the small width of the corresponding ``out'' correlation function
in Fig.\ 5 (b), while
the similar transverse size in all three cases leads to rather similar
``side'' correlation functions.

\begin{figure}\hspace*{0.6in} \vspace*{2cm}
\psfig{figure=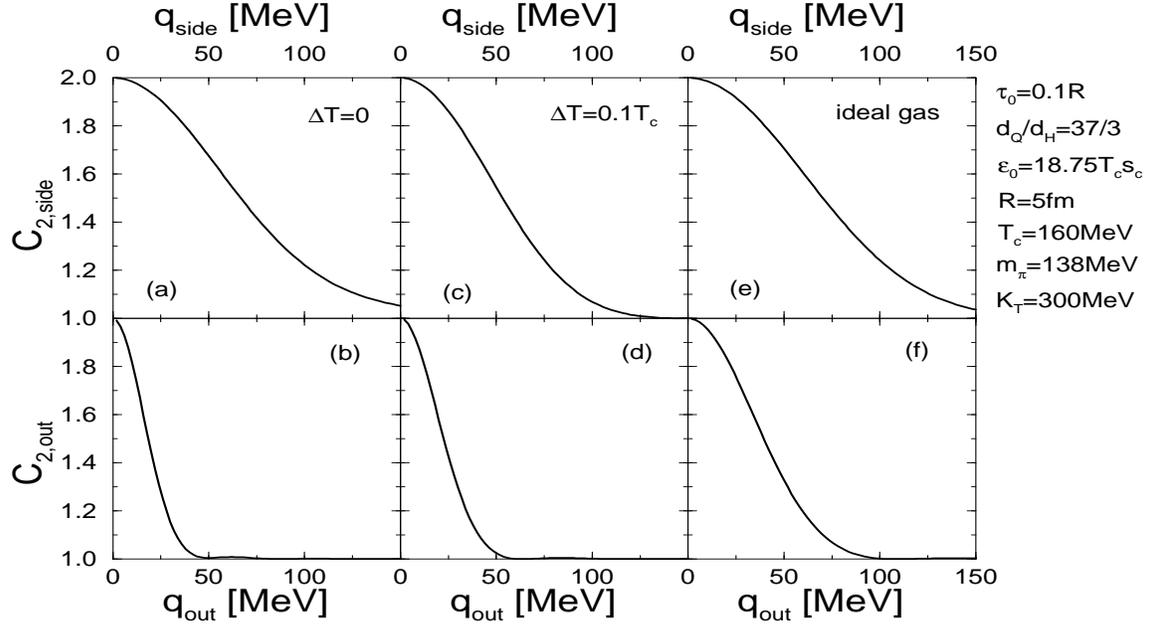,width=3in,height=3.5in,angle=-90}
\vspace*{-3cm}
\caption{``Side'' (a,c,e) and ``out'' correlation functions (b,d,f) 
calculated along the $T=0.7\, T_c$ isotherms of Figs.\ 3 (b,d,f).}
\end{figure}

Given the correlation functions, one then determines the 
ratio of inverse widths, $R_{\rm \,out} /$ $R_{\rm \,side}$. (The width
of the correlation function is here taken as the $q$--value
where $C_2 = 1.5$.) If the systems have similar transverse size, 
one then expects that this ratio is a good measure for the lifetime
of the system.
Figs.\ 4 (c,d) show $R_{\rm \,out} / R_{\rm \,side}$
as a function of the initial
energy density $\epsilon_0$. As expected, this ratio mirrors closely 
the dependence of the lifetime
on initial conditions in Figs.\ 4 (a,b). The effect is maximized around
initial energy densities expected to be reached at the RHIC collider 
\cite{glue}. The enhancement over the ideal gas case is of the order
of 40--100\% (for $\Delta T=0.1\, T_c$ to $\Delta T=0$).

At energy densities
estimated to be reached in CERN SPS Pb+Pb collisions ($\epsilon_0 \sim
1-2\, T_c\, s_c$ in our units), one expects from the above
that $R_{\rm \,out}/R_{\rm \,side} \sim 1.5 - 2$. 
However, present data from CERN SPS \cite{na44} 
indicate that the (fitted) out--radii are rather similar 
to the side--radii. This does not contradict my results, because, as 
shown by Schlei et al.\ \cite{schlei2} in the framework of a hydrodynamical
calculation similar to mine, 
correlation functions constructed from thermal pions only
give $R_{\rm \,out}/R_{\rm \,side} \sim 2$ (cf.\ especially 
\cite{schlei}), while the incorporation of long-lived
{\em resonance decays\/} leads to a reduction of that ratio and
good agreement with the measured radii.
Note that kaon interferometry \cite{miklos,schlei3,future}
is preferable, though experimentally more difficult,
because only distortions of the interference pattern
due to shorter lived $K^*$ resonances have to be taken into account.

\section{DIS- AND REAPPEARANCE OF TRANSVERSE DIRECTED FLOW}

The softening of the EoS and the delay in the expansion
have an interesting consequence for semi-peripheral heavy-ion collisions
at AGS energies. If the hot, compressed (baryon-rich) matter in the 
central zone undergoes a transition
to the QGP, its tendency to expand is reduced, similarly as discussed
above. This prevents the deflection of spectator matter, as it would occur
for a ``stiff'' EoS with a stronger tendency to
expand, for instance a purely hadronic EoS without
phase transition \cite{puersuen}. As shown in Fig.\ 6,
this effect is observable in the excitation function of the transverse 
directed flow per baryon,
\begin{equation}
\langle p_x/N \rangle^{dir} = \frac{1}{N} \int_{-y_{CM}}^{y_{CM}} {\rm d}y~
\langle p_x/N \rangle (y)~\frac{{\rm d}N}{{\rm d}y}~{\rm sgn} (y)~.
\end{equation}
The average transverse momentum per nucleon for
a given (fluid) rapidity $y$ is here defined as
$\langle p_x/N \rangle (y) = m_N\, \langle v_x \rangle \, \langle
\gamma \rangle$, where $m_N = 938$ MeV, $\langle \gamma \rangle \equiv
(1-\sum_i \langle v_i \rangle^2)^{-1/2}$, and
\begin{equation}
\langle v_i \rangle \equiv \frac{1}{{\rm \Delta}N} \,
\int_{-y-\Delta y/2}^{y+\Delta y/2} {\rm d} y\, v_i \,
\frac{{\rm d}N}{{\rm d}y}\,\, , \, \, \, i=x,y,z\,\,,
\end{equation}
$\Delta N = \int_{-y-\Delta y/2}^{y+\Delta y/2} {\rm d} y\,{\rm d}N/
{\rm d}y$.

\begin{figure}\hspace*{0.3in} \vspace*{0.5cm}
\psfig{figure=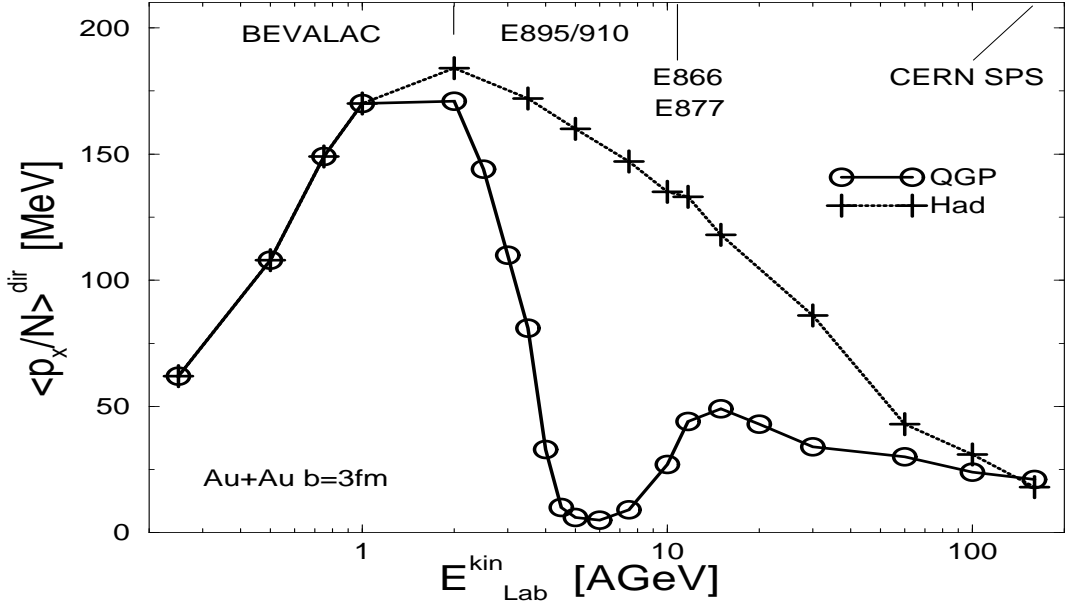,width=3.5in,height=3.5in,angle=-90}
\vspace*{-1.5cm}
\caption{The excitation function of the transverse directed flow as 
calculated from 3+1 dimensional
hydrodynamics for Au+Au collisions at impact parameter $b=3$ fm.
The full line (and the circles) is for an EoS with phase transition
to the QGP, the dotted curve (and the crosses) is for a purely hadronic
EoS.}
\end{figure}

The overall decrease of this quantity above $E_{\rm Lab}^{\rm kin} \sim
2$ AGeV observed for both EoS is 
simply due to the fact that faster spectators
are less easily deflected by the hot, expanding participant matter.
One clearly observes a dramatic {\em drop\/}
between BEVALAC and AGS beam energies  and
an {\em increase\/} beyond $\sim 10$ AGeV for the EoS with
phase transition
as compared to the calculation with the pure hadronic EoS. 
Thus, {\em there is a local minimum in the excitation function of the
directed transverse (in-reaction-plane) collective flow around $\sim 6$\/}
AGeV, which is again related to the phase transition to the QGP
and the existence of a ``soft region'' in the nuclear matter EoS. 
Note that the position of the minimum strongly depends on the details
of the EoS.
It may easily shift to higher beam energies, if more resonances are
included in the hadronic part of the EoS. Also, 
absolute values for the directed flow cannot yet be compared to 
experimentally measured ones, since at this stage freeze-out 
has not been performed. Moreover, (physical) viscosity is
neglected in the ideal hydrodynamic picture, which is known to
have a strong influence on flow \cite{schmidt}. On the other hand,
there is a certain amount of numerical viscosity inherent
in the transport scheme that solves the hydrodynamical equations.
This viscosity must be large enough to
suppress numerical instabilities \cite{test1,test2}, 
but it will in turn also affect the position of the minimum \cite{future2}.
Finally, the nature of the transition at {\em finite\/} baryon density
is completely unknown. In the above calculation, a sharp first order
transition was assumed \cite{puersuen}. A smooth transition
(similar to that in Fig.\ 1) would certainly tend to wash out the
minimum in the excitation function of the flow, as pressure
gradients become bigger and increase the system's tendency to expand
and deflect spectator matter.
The main point is, however, that irrespective of these quantitative
uncertainties, the {\em minimum\/} is a generic {\em qualitative\/}
signal for a transition from hadron
to quark and gluon degrees of freedom in the nuclear matter EoS.

\begin{figure}\hspace*{0.3in} 
\psfig{figure=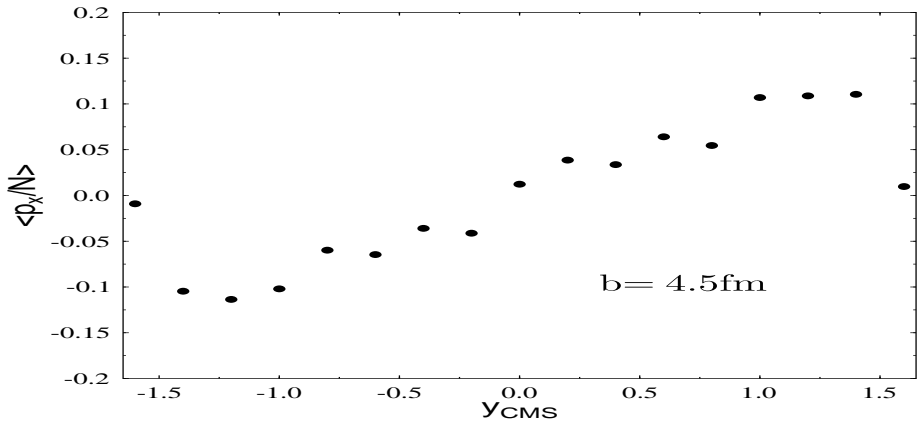,width=5in,height=4in}
\vspace*{-1.5cm}
\caption{The transverse momentum as a function of rapidity as calculated 
from 3+1 dimensional three--fluid dynamics for a Pb+Pb collision
at 11 AGeV and impact parameter $b=4.5$ fm.}
\end{figure}

A more realistic description of heavy-ion collisions, especially
for higher bombarding energies, can
be achieved with the so-called three-fluid model \cite{3f}. Ordinary one-fluid 
dynamics assumes instantaneous local equilibrium, even in the initial
collision stage. This leads to instantaneous stopping of target
and projectile, which is obviously unrealistic since
the stopping power of nuclear matter is finite. In the three-fluid
approach, one solves this problem by assuming that target and projectile 
are separate fluids which
interact via collision terms derived from kinetic theory. Collisions
among target and projectile fluid elements create a third fluid.
Fig.\ 7 shows a calculation of $\langle p_x/N \rangle (y)$ for
a semi-peripheral ($b=4.5$ fm) Pb+Pb collision at 11 AGeV. 
Since this is an ongoing investigation, I am yet not able to show results
for the excitation function of the transverse directed flow.

\section{CONCLUSIONS}

In this paper I have discussed the softening of the nuclear matter
EoS due to a transition to the QGP. Since 
present lattice data only constrain the width of
the transition region to be in the range 
$0\leq \Delta T < 0.1\, T_c$, it is important
to test how such uncertainties may influence dynamical observables.
This has to be investigated in the framework of
relativistic hydrodynamics, since that is the only model that provides
a direct link between the EoS and collective observables.

I focussed on the lifetime of the system as a function of
initial energy density as one possible collective observable 
that can discriminate between different EoS.
One--dimensional studies in the framework of
the Landau expansion model \cite{dhrmg} show
that the lifetime is much longer
in the case of a first order phase transition, $\Delta T=0$,
as compared to the expansion of an ideal gas without transition. The
prolongation of the lifetime can be up to a factor of 10,
provided the initial energy density corresponds to that of mixed phase
with a large fraction of QGP. In that case, rarefaction proceeds
as a (slow) deflagration shock wave instead of a simple rarefaction
wave (with the speed of sound as characteristic propagation
velocity). In the case of a smooth transition with $\Delta T= 0.1\, T_c$,
deflagration solutions no longer exist, but due to the reduction
of the velocity of sound in the transition region the expansion
is still slower than that of an ideal gas. Thus, the lifetimes
(of matter with $T=0.7\, T_c$) remain on the order of
a factor of 7 longer as compared to the ideal gas expansion.

The results for the Bjorken cylinder expansion \cite{dhrmg2} 
were similar as for the 
Landau expansion, up to two important exceptions:
(a) the maximum lifetimes emerged at higher initial energy densities
corresponding to QGP matter instead of mixed phase matter, 
and (b) the lifetimes were in general shorter.
Both effects are explained by the very efficient cooling due to the
initial longitudinal velocity profile associated with the boost
invariance of the problem. This effect causes an overall
reduction of the lifetimes. Moreover, in order to remain long enough
in the soft region of the EoS and thus
to slow down (or even stall) the cooling process in transverse direction, 
one has to start at higher initial energy 
densities to compensate for the longitudinal cooling. Otherwise, 
the longitudinal cooling reduces the energy densities too fast and 
one leaves the soft region of the EoS too soon. Then, the expansion and
cooling is accelerated due to the higher velocity of sound outside the
soft region of the EoS.

I then showed (cf.\ also \cite{pratt})
that the prolongation of the lifetime can be
observable via the ratio $R_{\rm \,out}/R_{\rm \,side}$ of inverse
widths of two--particle correlation functions in out-- and side--direction.
This ratio follows the behaviour of the lifetimes rather closely. The 
prolongation of the lifetime in the case of a transition to the
QGP could therefore be in principle searched for using 
this observable. The enhancement of that ratio
is, of course, strongest in the case that the transition is first order with
a large latent heat. An interesting result is that, for 
the Bjorken cylinder geometry,
the maximum of $R_{\rm \,out}/R_{\rm \,side}$ occurs at energy densities
presumably reached at the RHIC collider. 

There are several effects which may reduce the strength of the time--delay
signal observable via the $R_{\rm \,out}/R_{\rm \,side}$ ratios
that will require further investigation.
First, the decay of long-lived resonances can simulate
time delay \cite{padula}.
Interferometry with kaons instead of pions is therefore preferable
\cite{future}. Finally, while the present investigations covered a wide
range of uncertainties
in the EoS, the calculations have neglected effects of
dissipation that tend in general to reduce
the collective flow strengths predicted via ideal hydrodynamics.
For instance, bulk viscosity appears
in the hydrodynamical equations of motion in a similar way as the
pressure, and could in principle counteract any reduction of the
velocity of sound in the transition region. 
The main result of this paper is, nevertheless, that the generic
time--delay signature of QGP formation is remarkable robust to present
uncertainties in the QCD equation of state.

I finally discussed the dis-- and reappearance of the transverse directed
flow around AGS energies. As a result of the
softening of the EoS in the transition region, the tendency for
the hot and dense reaction zone to expand is considerably reduced.
Thus, spectator matter passes this zone undeflected, which consequently leads
to a substantial decrease in the transverse flow as compared to a
purely hadronic scenario without transition to the QGP.
Quantitative estimates for the size of this effect are subject to
large uncertainties,
as it appears to be sensitive to the EoS used, 
and is influenced by numerical and (so far neglected) physical viscosity.
Also, freeze-out has not yet been performed to calculate the
particle (instead of fluid) flow. The local minimum in the excitation 
function of the transverse directed flow remains, however, an important
{\em qualitative\/} signal for a transition (to the QGP) in the nuclear 
matter EoS.
\newpage
\noindent
{\bf Acknowledgments}
\\ ~~ \\
I would like to thank the organizers for their invitation to present
these results at Quark Matter '96.
I am indebted to G.\ Bertsch, U.\ Heinz, D.\ Keane, B.\ Schlei,
E.\ Shuryak, and W.\ Zajc for valuable comments and discussions on
time delay and interferometry.
Special thanks go to J.\ Brachmann and A.\ Dumitru for their efforts to
provide me with Figure 7, and to S.\ Bernard for discussions
related to 1+1 dimensional transport algorithms. Finally, I am most grateful
to Miklos Gyulassy for his continuing interest and his encouragement 
to pursue this subject, for discussions, and for all other kinds of
support during my time at Columbia University.


\begin{thebibliography}{99}
\bibitem{footnote1}
Renewed interest in that signature,
has, however, emerged with the advent of recent results from
the NA50 experiment at the CERN SPS, showing a rapid variation 
of $J/\Psi$--suppression as a function of transverse energy in
Pb+Pb collisions at 158 AGeV,
see the contribution of M.\ Gonin in these proceedings.
\end{thebibliography}
\end{document}